\documentclass[twocolumn]{revtex4-1}

\usepackage{graphicx}
\usepackage[utf8]{inputenc} 
\usepackage{bm}
\usepackage{amsmath}
\usepackage{amssymb}
\usepackage[normalem]{ulem}

\begin{document}
	
\title{Determining phoretic mobilities with Onsager's reciprocal relations: electro- and thermophoresis revisited}
	
\author{J\'er\^ome Burelbach, Holger Stark}
	
\address{Institut f\"ur Theoretische Physik, Technische Universit\"at Berlin, Hardenbergstraße 36, 10623 Berlin, Germany}

\date{October 2018}
	
\maketitle

\section{Abstract}
We use a hydrodynamic reciprocal approach to phoretic motion to derive general expressions for the electrophoretic and thermophoretic mobility of weakly charged colloids in aqueous electrolyte solutions. Our approach shows that phoretic motion can be understood in terms of the interfacial transport of thermodynamic excess quantities that arises when a colloid is kept stationary inside a bulk fluid flow. The obtained expressions for the mobilities are extensions of previously known results as they can account for different hydrodynamic boundary conditions at the colloidal surface, irrespective of how the colloid-fluid interaction range compares to the colloidal radius.

\section{Introduction}

Phoresis is the directed motion of colloids through a fluid in response to a thermodynamic gradient. Despite many theoretical advances \cite{smoluchowski1903contribution,huckel1924cataphoresis,Derjaguin1966,Ruckenstein1981,Agar1989,Anderson1989}, phoretic motion remains a fruitful subject for current research \cite{Parola2004,Ajdari2006,Wurger2007,Dhont2007,Khair2009,Brady2011,Semenov2015}. The majority of existing theories make use of the fact that phoretic motion is a force-free interfacial phenomenon: It relies on the presence of a specific colloid-fluid interaction and is accompanied by an interfacial fluid flow in the opposite direction. This interplay between colloidal motion and interfacial fluid flows suggests that phoresis has a hydrodynamic character that cannot be derived from a purely equilibrium-thermodynamic consideration. 

Indeed, multiple treatments of different phoretic phenomena based on force-free arguments have repeatedly shown that the driving force of phoretic motion should depend on the hydrodynamic boundary condition at the colloidal surface \cite{Ajdari2006,Khair2009,Morthomas2009,Gaspard2018}. However, the argumentation of most force-free approaches is restricted to the case where the colloid-fluid interaction range is very short compared to the colloidal radius, a case also known as the boundary layer approximation (BLA). Although attempts have been made to generalise the force-free argument to cases beyond the BLA \cite{Piazza2004,Dhont2008}, these works only considered a stick boundary condition at the colloidal surface and yielded different predictions for the size-dependence of colloidal thermophoresis. 

In order to solve this issue, a unified description of phoretic motion within the framework of non-equilibrium thermodynamics (NET) has recently been proposed by Burelbach $et$ $al.$ \cite{Burelbach2018b}. In this description, a link between the hydrodynamic and thermodynamic approaches to phoresis was drawn, and it was shown that Onsager's reciprocal relations can be used to determine the colloidal drift velocity for any hydrodynamic boundary condition, irrespective of how the interaction range compares to the colloidal size. Here, we apply this approach to the phoretic motion of weakly charged colloids in aqueous electrolyte solutions, to obtain generalised expressions for the electrophoretic and thermophoretic mobility. As we are interested in the interfacially-driven phoretic motion of single colloids, gravitational sedimentation, Brownian motion and pair-interactions of the colloids will be ignored in the following discussion. We start by elaborating on the hydrodynamic reciprocal theory introduced in \cite{Burelbach2018b}.

\section{Onsager's reciprocal relations}

\subsection{General formalism}
According to Onsager's theory of NET, the heat flux $\mathbf{J}_{q}$ and particle fluxes $\mathbf{J}_{i}$ inside an $N$-component system are described by phenomenological expressions that are linear in the thermodynamic gradients \cite{deGroot1962}
\begin{eqnarray}
\mathbf{J}_{i} & = & L_{iq}\nabla\frac{1}{T}+\frac{1}{T}\sum_{k}L_{ik}\left( -\nabla_T\mu_{k}+\mathbf{F}_{k}\right), \label{eq:th-10}\\
\mathbf{J}_{q} & = & L_{qq}\nabla\frac{1}{T}+\frac{1}{T}\sum_{k}L_{qk}\left( -\nabla_T\mu_{k}+\mathbf{F}_{k}\right), \label{eq:th-11}
\end{eqnarray}
where $T$ is the temperature, $\mu_k$ is the chemical potential of component $k$ and $\mathbf F_k$ is the body force on component $k$. The notation $\nabla_T$ means that the gradient is evaluated at constant temperature. The coefficients $L$ are known as the Onsager transport coefficients. Here, we are interested in the motion of colloids suspended in a fluid that consists of solvent ($e.g.$ water) and small solutes ($e.g.$ ions). In the following, we use the indices $i=1$ for the colloids, $i=0$ for the solvent and $i>1$ for the solutes. 

The colloidal flux induced by a body force $\mathbf{F}_1$ on the colloids can be expressed as $\mathbf{J}_1=c\mathbf{F}_1/\xi$, where $c$ is the colloidal concentration and $\xi$ is the friction coefficient of a colloid. Thus, the Onsager coefficient $L_{11}$ is related to the friction coefficient via
\begin{equation}
L_{11} = \frac{cT}{\xi}.\label{eq:-0}
\end{equation}  
The friction coefficient is given by $\xi=6\pi b\eta R$, where $\eta$ is the fluid viscosity. The slip coefficient $b=(1+2l_s/R)/(1+3l_s/R)$ depends on the ratio $l_s/R$ between the slip length $l_s$ and the radius $R$ of the colloid \cite{Barber}. As a result, $b$ takes the value $b=1$ for a stick boundary condition and $b=2/3$ for a perfect-slip boundary condition at the colloidal surface.

As an interfacial phenomenon, phoretic motion relies on a specific interaction between the colloid and the fluid, leading to an interfacial excess of fluid particles (and fluid enthalpy) at the colloidal surface. This excess is located inside a region also known as the interfacial layer, whose effective width $\lambda$ quantifies the range and steepness of the interaction potential. The layer is termed 'thin' if the potential decays rapidly over a distance small compared to the colloidal radius ($R/\lambda\gg 1$), and 'wide' otherwise ($R/\lambda\ll 1$). The relevant interfacial part of the chemical potential of the colloid is related to the interfacial tension $\gamma$ via $\mu_1=A_c\gamma$, where $A_c$ is the constant surface area of the colloid. The interfacial chemical potential can be related to excess quantities of the fluid via the Gibbs adsorption equation \cite{Burelbach2018b}:
\begin{equation}
-d\mu_1=H_\phi\frac{dT}{T}+\sum_{k\neq 1}N_k^\phi (d\mu_k)_T,\label{eq:-26}
\end{equation}
where $N_{k}^{\phi}$ is the net interfacial excess of fluid component $k$ and $H_{\phi}$ is the interfacial excess enthalpy. The gradient of $\mu_1$ at constant temperature can hence be expressed as 
\begin{equation}
-\nabla_T\mu_1=\sum_{k\neq 1}N_k^\phi \nabla_T\mu_k.\label{eq:-29}
\end{equation}
The condition of charge neutrality further implies that the (electric) body forces on the colloid and its interfacial layer cancel:
\begin{equation}
\mathbf F_1=-\sum_{k\neq 1}N_k^\phi\mathbf F_k.\label{eq:-30}
\end{equation}
Although phoretic motion is a force-free phenomenon, an effective phoretic driving force $\mathbf F_{\text{ph}}$ can be defined as the equal and opposite of the external force that needs to be exerted on the colloid to cancel its phoretic velocity $\mathbf v_{\text{ph}}=\mathbf J_1/c$, such that $\mathbf F_{\text{ph}}= \xi\mathbf v_{\text{ph}}$. By convention, phoretic motion is commonly described in terms of fluid degrees of freedom only, without reference to the equation of state of the colloidal component \cite{Derjaguin1987,Anderson1989}. To this end, the colloidal degrees of freedom can be eliminated in eq. (\ref{eq:th-10}) for $\mathbf J_1$ by using eqs. (\ref{eq:-29}) and (\ref{eq:-30}). The phoretic force $\mathbf F_{\text{ph}}= \xi\mathbf J_1/c$ thus takes the linear form
\begin{equation}
\mathbf F_{\text{ph}}=-Q^{*}\frac{\nabla T}{T}+\sum_{k\neq1}N_k^{*}\left( -\nabla_T\mu_{k}+\mathbf{F}_{k}\right),\label{eq:-73}
\end{equation} 
where, in view of eq. (\ref{eq:-0}), the coefficients $Q^{*}$ and $N_k^{*}$ are related to the Onsager coefficients of the colloids via

\begin{equation}
Q^{*}=\frac{L_{1q}}{L_{11}}\hspace{0.3cm}\text{and}\hspace{0.3cm}N_k^{*} =\frac{L_{1k}}{L_{11}}-N_{k}^{\phi}.\label{eq:-24}
\end{equation} 

From eq. (\ref{eq:-73}), we see that the phoretic force consists of two different contributions: a 'thermal' term that couples to the temperature gradient $\nabla T$ and an 'electrochemical' term coupling to the electrochemical force $-\nabla_T\mu_k+\mathbf F_k$ on the fluid components (solvent and solutes).

\subsection{The reciprocal argument for phoretic motion}

The phoretic drift velocity of a colloid is usually determined based on momentum conservation in a force-free system. However, the application of force-free arguments is only straightforward when the interaction range between the colloid and the fluid is either very small or very large compared to the colloidal size \cite{smoluchowski1903contribution,huckel1924cataphoresis,Fayolle2008,Morthomas2008}. To overcome this limitation, we use an alternative hydrodynamic approach based on Onsager's reciprocal relations \cite{Onsager1931,Onsager1931a}
\begin{eqnarray}
L_{1k}=L_{k1}\hspace{0.3cm}\text{and}\hspace{0.3cm}L_{1q}=L_{q1}\label{eq:-1}.
\end{eqnarray}
These relations show that the colloidal flux couples to a temperature gradient or electrochemical force in the same way as heat and fluid particle fluxes couple to an external force on the colloid. As a consequence, the coefficients $Q^{*}$ and $N_k^{*}$ correspond to the interfacial transport of heat and fluid particles that arises when a colloid is subjected to an external force inside a homogeneous fluid at uniform temperature. 

Therefore, let us consider a single colloid of radius $R$ inside an infinitely extended, homogeneous fluid at uniform temperature. The bulk fluid (solvent and solutes) is moving at a uniform flow velocity $\mathbf u_\infty=u_\infty\mathbf{\hat y}$ and the colloid is kept at rest by an external force $\mathbf F_1=-\xi \mathbf u_\infty$. As the fluid flows through the interfacial layer, it carries an excess of heat and fluid particles into the bulk. For the interfacial layer to remain in a local thermodynamic equilibrium, the excess carried out on one side must be balanced by an equal influx of heat/fluid particles from the bulk on the other side of the layer. The volume-integrated fluxes of heat and fluid particles resulting from this interfacial exchange of fluid between the colloid and the bulk can be expressed as \cite{Agar1989,Burelbach2018b,Burelbachthesis}
\begin{eqnarray}
\mathbf J_{q,V} & = & \int_{R}^{\infty}h_\phi(r)\mathbf u\left(\mathbf r \right)dV,\label{eq:-23}\\
\mathbf J_{k,V} & = & \int_{R}^{\infty}n_k^\phi(r)\left( \mathbf u\left(\mathbf r \right)-\mathbf u_\infty\right) dV\label{eq:-5},
\end{eqnarray}
where $\mathbf u$ is the fluid flow velocity and $\mathbf{r}=r\mathbf{\hat r}$ is the position with respect to the centre of the colloid. $n_k^\phi$ is the interfacial excess number density of fluid component $k$ and $h_\phi$ is the interfacial excess enthalpy density. Particle fluxes are computed with respect to the bulk velocity $\mathbf u_\infty$ \cite{deGroot1962}, which has therefore been subtracted from the flow velocity in eq. (\ref{eq:-5}). 
If the solvent is incompressible ($n_0^\phi=0$), eq. (\ref{eq:-5}) only refers to the fluxes of solute particles inside the system. In view of eqs. (\ref{eq:th-10}) and (\ref{eq:th-11}), the Onsager forms of these volume-integrated fluxes are simply given by
\begin{equation}
\mathbf J_{q,V} = V\frac{L_{q1}}{T}\mathbf F_1,\hspace{0.3 cm}
\mathbf J_{k,V} = V\frac{L_{k1}}{T}\mathbf F_1,\label{eq:-19}
\end{equation}
where $V$ is the volume of the system. Using Onsager's reciprocal relations (\ref{eq:-1}) and the relation $\mathbf F_1=-\xi \mathbf u_\infty$ in eq. (\ref{eq:-19}), we obtain
\begin{eqnarray}
\mathbf J_{q,V} & = & -\frac{L_{1q}}{L_{11}}\mathbf u_\infty = -Q^*\mathbf u_\infty,\label{eq:-9}\\
\mathbf J_{k,V} & = & -\frac{L_{k1}}{L_{11}}\mathbf u_\infty = -\left( N_k^*+N_k^\phi\right) \mathbf u_\infty.\label{eq:-10}
\end{eqnarray}
For convenience, let us denote the interfacial excess densities $h_{\phi}$ and $n_{k}^{\phi}$ by $x_{\phi}$ and the corresponding interfacial transport coefficients $Q^*$ and $N_k^*$ by $X^*$. By respectively combining eqs. (\ref{eq:-23}) and (\ref{eq:-5}) with eqs. (\ref{eq:-9}) and (\ref{eq:-10}), and by using eq. (\ref{eq:-24}), we find the general transport relation
\begin{equation}
X^*\mathbf u_\infty=-\int_{R}^{\infty}x_{\phi}(r)\mathbf u\left(\mathbf r \right)dV.\label{eq:th-4}
\end{equation}
At local thermodynamic equilibrium, the interfacial layer remains unperturbed by the fluid flows and therefore spherically symmetric \cite{Agar1989}. As a result, the excess density $x_{\phi}(r)$ only depends on the distance $r$ from the colloidal centre \cite{Burelbach2018b}. Due to the circular symmetry along the direction $\mathbf{\hat y}$ of the bulk flow, only the $y$-component of $\mathbf u$ contributes to the volume integral. The angular integration in eq. (\ref{eq:th-4}) can hence be carried out separately over $\mathbf u$, giving $X^*u_\infty= -\int_{R}^{\infty}x_{\phi}(r)\left\langle \mathbf u\left(\mathbf r \right)\cdot\mathbf {\hat y}\right\rangle 4\pi r^2dr$. Here, the orientational average is $\left\langle \mathbf u\left(\mathbf r \right)\cdot\mathbf {\hat y}\right\rangle=\frac{1}{2}\int_{0}^{\pi}u_y(r,\theta)\sin\theta d\theta$, where $u_y(r,\theta)$ is the $y$-component of $\mathbf u$ and $\theta$ is the angle with respect to the $\mathbf{\hat y}$-axis. In view of eq. (\ref{eq:th-4}), an analytical solution of the flow field is required to obtain an explicit expression for the interfacial transport coefficient $X^*$. The key advantage of our reciprocal approach is that the flow velocity $\mathbf u$ around a stationary colloid has a well-known analytical solution \cite{Barber,Landau1987}, with spherical coordinates
\begin{eqnarray}
u_r & = & u_\infty\cos\theta \left\{1-\frac{3b}{2}\frac{R}{r}+\left(\frac{3b}{2}-1\right)\left(\frac{R}{r}\right)^3\right\}\nonumber\\ 
u_\theta & = & -u_\infty\sin\theta \left\{1-\frac{3b}{4}\frac{R}{r}-\frac{1}{2}\left(\frac{3b}{2}-1\right)\left(\frac{R}{r}\right)^3\right\}.\nonumber
\end{eqnarray}
For $b=1$, this flow field just reduces to the usual Stokes flow around a stationary sphere with a stick boundary condition. The orientational average can now be evaluated in a straightforward manner, yielding
\begin{equation}
X^*[x_\phi]=-\int_{R}^{\infty}x_{\phi}(r)\left(1-b\frac{R}{r}\right)4\pi r^2dr,\label{eq:th-54}
\end{equation}
where the notation $X^*[x_\phi]$ means that the transport coefficient $X^*$ is a functional of $x_\phi$. 

Eq. (\ref{eq:th-54}) constitutes the general hydrodynamic form of the interfacial transport coefficient $X^*$ as presented in \cite{Burelbach2018b} and completely determines the phoretic force $\mathbf F_{\text{ph}}$ given by eq. (\ref{eq:-73}). Eq. (\ref{eq:th-54}) shows that the presence of a solid surface leads to viscous forces that tend to reduce the strength of phoretic motion (due to the term $-bR/r$), and that these viscous forces are stronger for stick ($b=1$) than for perfect slip ($b=2/3$). The validity of eq. (\ref{eq:th-54}) and its dependence on hydrodynamic boundary conditions have recently been confirmed by means of computer simulations \cite{Burelbach2018}. For a stick boundary condition, the form of eq. (\ref{eq:th-54}) coincides with the thermophoretic force in \cite{Parola2004}, which used a different non-conservative term instead of the interfacial excess enthalpy density. One limitation of eq. (\ref{eq:th-54}) is that it only holds if the applied thermodynamic gradients are not strongly modified by the colloids.

In the point-like limit ($R\rightarrow0$, or $b=0$), the viscous term vanishes and eq. (\ref{eq:th-54}) just corresponds to the net excess of fluid particles $N_k^\phi=\int n_k^\phi dV$ and enthalpy $H_\phi=\int h_\phi dV$ inside the interfacial layer. Using eqs. (\ref{eq:-26}) and (\ref{eq:-30}) in eq. (\ref{eq:-73}), the phoretic force can then alternatively be expressed as \begin{equation}
\mathbf F_{\text{ph}}=-\nabla\mu_1+\mathbf F_1,\hspace{0.4 cm}R\rightarrow0,\label{eq:-32}
\end{equation}
showing that the point-like limit corresponds to a thermodynamic treatment of phoretic motion that ignores the hydrodynamic boundary condition at the particle surface. 

For thin layers ($R\gg\lambda$), a leading order expansion of eq. (\ref{eq:th-54}) in the small parameter $z=r-R$ allows the recovery of the well-known BLA result for a stick boundary condition ($b=1$), given by the Smoluchowski-Derjaguin integral \cite{Derjaguin1987,smoluchowski1903contribution}:
\begin{equation}
X^*[x_\phi]=-4\pi R\int_{0}^{\infty}zx_{\phi}(z)dz,\hspace{0.4 cm}R\gg\lambda.\label{eq:-31}
\end{equation}
To leading order in $R/\lambda\gg 1$, eq. (\ref{eq:th-54}) also complies with the forms of the diffusiophoretic and thermophoretic mobilities respectively obtained in \cite{Gaspard2018,Morthomas2009} using the BLA. 

The remainder of this letter will focus on the evaluation of transport coefficients for charged colloids undergoing electro- or thermophoresis in aqueous electrolyte solutions. We consider the case of weakly charged colloids in order to obtain a general analytical solution for the electric potential around a charged sphere, which is required for the evaluation of eq. (\ref{eq:th-54}).

\section{Phoretic motion in aqueous suspensions}

\subsection{Electrophoresis}

The system of interest is a dilute aqueous suspension of weakly charged colloids ($i=1$), immersed in a fluid that consists of water ($i=0$, assumed incompressible) and multiple ionic solutes ($i>1$, ions). The interaction of the water molecules with the ions and the charged colloids leads to the formation of hydration layers around these charged species. The electrostatic interaction between the colloids and the ions leads to the build-up of electric double layers around the colloids. The ions are treated as point-like particles that follow a Poisson-Boltzmann distribution inside the electric double layer. We denote the electric potential of the colloid by $\varphi$ and the potential energy of a double-layer ion as $\phi_k=q_k\varphi$, where $q_k$ is the corresponding charge. 

Within the Debye-H\"uckel approximation for weakly charged colloids, the potential energy of an ion is much smaller than the thermal energy, so that the local ion density $n_k=n_k^b\exp{\left[-\phi_k/(k_BT)\right]}$ can be expanded to quadratic order in the small parameter $\phi_k'=\phi_k/(k_BT)\ll 1$, where $n_k^b$ is the bulk ion density. The interfacial excess density $n_k^\phi$ required for the evaluation of $N_k^*$ can then be written as $n_k^\phi=n_k^b\left(-\phi_k'+\phi_k'^2/2\right)$ and the electric potential of the colloid obeys the linearised Poisson equation $\nabla^2\varphi=\kappa^2\varphi$, where $\kappa=\left[\left(\sum_k n_k^bq_k^2\right)/(\epsilon k_BT)\right]^{\frac{1}{2}}$ is the inverse of the Debye screening length $\lambda$. The solution is given by the well-known Yukawa potential
\begin{equation}
\varphi(r)=\zeta\frac{R}{r}\exp{-\kappa(r-R)},\label{eq:-11}
\end{equation}
where 
\begin{equation}
\zeta=\varphi(R)=\frac{\sigma R}{\epsilon(1+\kappa R)}\label{eq:-18}
\end{equation}
is the electric surface potential, $\epsilon$ is the electric permittivity, and $\sigma$ is the surface charge density of the colloid. 

By definition, electrophoresis occurs when the system is exclusively subjected to an electric field $\mathbf E$. The applied electric field is usually reversed periodically, as to avoid the induction of ionic companion fields inside the fluid. The colloids acquire a phoretic velocity $\mathbf v_{\text{ph}}=\mu_E\mathbf E$, where $\mu_E$ is the electrophoretic mobility, and eq. (\ref{eq:-73}) simply reduces to $\mathbf F_{\text{ph}}=\sum_{k}N_k^{*}q_k\mathbf{E}$. With $\mathbf F_{\text{ph}}= \xi\mathbf v_{\text{ph}}$, the electrophoretic mobility is thus related to the interfacial transport of charges via 
\begin{equation}
\mu_E=\frac{1}{\xi}\sum_{k\in \text{ions}}N_k^*q_k\label{eq:-4}.
\end{equation}

In view of eqs. (\ref{eq:th-54}) and (\ref{eq:-11}), and for later use, it is convenient to introduce the integral form
\begin{equation}
\mathcal{I}^{m}_n=\int_{1}^{\infty}\left(1-b\rho^{-1}\right)\rho^{1-m}\exp{\left\{-n\kappa'(\rho-1)\right\}}d\rho, \label{eq:-12}
\end{equation}
where $\rho=r/R$ and $\kappa'=\kappa R$. Using eqs. (\ref{eq:th-54}) to evaluate eq. (\ref{eq:-4}), the charge transport can then simply be expressed as $\sum_k N_k^*q_k=4\pi R\epsilon\zeta\kappa'^2\mathcal I^0_1$, where $\mathcal I^0_1=\\ \left(1+\kappa'(1-b)\right)/\kappa'^2$. Combining this with eq. (\ref{eq:-4}), we obtain an explicit expression for the electrophoretic mobility:
\begin{equation}
\mu_E=\frac{2}{3}\frac{\epsilon\zeta}{b\eta}\left(1+\kappa'(1-b)\right)\label{eq:-6}.
\end{equation}
This expression can be compared to the well-known limiting cases of 'wide layers' ($\kappa'\ll 1$) and 'thin layers' ($\kappa'\gg 1$). The former case is also sometimes referred to as the 'H\"uckel limit' \cite{huckel1924cataphoresis,Morthomas2008}, whereas the latter case corresponds to the 'Smoluchowski limit' (BLA) \cite{smoluchowski1903contribution,Anderson1989}. As $\kappa'$ is the ratio between the colloidal radius and the Debye length, we study the $\kappa'$-dependence of $\mu_E$ by fixing the colloidal size, so that $\kappa'$ only varies with ionic strength. A general expression of $\mu_E$ in terms of $\kappa'$ is obtained by substituting eq. (\ref{eq:-18}) into eq. (\ref{eq:-6}), giving

\begin{equation}
\mu_E=\frac{2\sigma R}{3b\eta}\frac{1+\kappa'(1-b)}{1+\kappa'}\label{eq:-22}.
\end{equation}

For wide layers ($\kappa'\ll 1$), the electrophoretic mobility reduces to $\mu_E=2\sigma R/(3b\eta)$, which for a stick boundary ($b=1$) coincides with the well-known 'H\"uckel' expression for electrophoresis \cite{huckel1924cataphoresis}. For thin layers ($\kappa'\gg 1$), a stick boundary condition ($b=1$) yields $\mu_E=2\sigma/(3\eta\kappa)$, which only differs from the 'Smoluchowski' expression for electrophoresis by a factor $2/3$ \cite{smoluchowski1903contribution}. This is because the Smoluchowski expression assumes the electric permittivity of the colloid to be negligible compared to that of the fluid, whereas we have assumed that they are the same. However, the electrophoretic mobility takes a quite different form in the limit $\kappa'\gg 1$ when there is hydrodynamic slip at the colloidal surface ($b\neq 1$):
\begin{equation}
\mu_E=\frac{2}{3}\frac{\sigma R}{b\eta}(1-b),\hspace{0.4 cm}\kappa'\gg 1\hspace{0.2 cm}\text{and}\hspace{0.2 cm}b\neq 1\label{eq:-7}.
\end{equation}
This expression reduces to the result for weakly charged colloids derived in \cite{Khair2009} when the slip length is small compared to the radius ($l_s/R\ll1$). Equation (\ref{eq:-7}) suggests that the electrophoretic mobility does not vanish with increasing ionic strength if there is slip at the colloidal surface. The reason is that electrophoresis relies on the ability of the fluid to flow inside the electric double layer. Such a flow is prevented by a stick boundary condition at high ionic strength, when the layer is completely 'squeezed' onto the colloidal surface. However, the fluid can flow inside the layer if there is hydrodynamic slip at the surface, thus allowing electrophoretic motion at high ionic strength.

\subsection{Thermophoresis}

The phoretic motion of colloids in a temperature gradient is known as thermophoresis \cite{Piazza2008}. The thermophoretic mobility $D_T$ is defined by
\begin{equation}
\mathbf v_{\text{ph}}=-D_T\nabla T\label{eq:-2}.
\end{equation}
Unlike electrophoresis, thermophoresis is often studied in the presence of a stationary gradient. The ions respond to the temperature gradient much faster than the colloids and can therefore be assumed at steady state while the colloids undergo thermophoretic motion \cite{Wurger2010}.

In view of eq. (\ref{eq:th-10}), the ionic solute flux takes the form $\mathbf{J}_{k}=L_{kq}\nabla(1/T)+L_{kk}/T\left( -\nabla_T\mu_{k}+\mathbf{F}_{k}\right)$. Here, we have used the fact that there is no excess of fluid particles around solute $k$, based on the assumption that water is incompressible ($n_0^\phi=0$) and that the solutes are point-like particles within the Poisson-Boltzmann mean field approximation. As a result, the ionic solute flux $\mathbf J_k$ does not couple to the chemical potential gradients of other fluid components. The electric force $\mathbf F_k=q_k\mathbf E_T$ on the ion derives from a thermoelectric field $\mathbf E_T$ in the bulk of the suspension \cite{Putnam2005}. This field is due to the accumulation of ions and counterions on opposite sides of the system and can be written as $\mathbf E_T=-\varphi_T\nabla T/T$, where $\varphi_T$ is the thermoelectric potential. The solute steady-state condition $\mathbf J_k=0$ yields
\begin{equation}
-\nabla_T\mu_k+\mathbf F_k=Q_{k}^*\frac{\nabla T}{T}.\label{eq:-21}
\end{equation}
The heat transport coefficient $Q_k^*=L_{kq}/L_{kk}$ of ionic solute $k$ comprises an interfacial part $Q_{k0}^{*}$, related to the enthalpy density of its hydration layer \cite{Agar1989}, and a thermal contribution $k_BT$ due to its Brownian motion \cite{Burelbachthesis}: 
\begin{equation}
Q_{k}^*=Q_{k0}^{*}+k_BT.
\end{equation} 
Moreover, the chemical potential gradient at constant temperature of a point-like solute is simply given by $\nabla_T\mu_k=k_BT\nabla\ln n_k^b$. Based on the condition of charge neutrality $\sum_k n_k^bq_k=0$, an explicit expression for the thermoelectric potential $\varphi_T$ can be obtained by multiplying eq. (\ref{eq:-21}) by $n_k^bq_k$ and summing over all ionic solutes:
\begin{equation}
\varphi_T=-\frac{\sum_k n_k^bq_kQ_{k0}^*}{\sum_k n_k^bq_k^2}.\label{eq:-27}
\end{equation}

Substituting eq. (\ref{eq:-21}) into eq. (\ref{eq:-73}), the thermophoretic force on the colloid finally takes the form $\mathbf F_{\text{ph}} = -\xi D_T\nabla T$, where the thermophoretic mobility can be identified as
\begin{equation}
D_{T}=\frac{1}{\xi T}\left(Q^*-\sum_{k\in \text{ions}}N_{k}^{*}\left( Q^*_{k0}+k_BT\right)\right).\label{eq:-8}
\end{equation}
The thermophoretic mobility of the colloid thus comprises a thermal term due to the interfacial heat transport coefficient $Q^*$ of the colloid, and an electrochemical term due to the interfacial solute transport coefficient $N_k^*$ of the colloid, which in turn couples to the heat transport coefficient $Q_k^*$ of solute $k$ at steady state. Given that $Q^*$ and $N_k^*$ have the same functional form $X^*$ as defined in eq. (\ref{eq:th-54}), eq. (\ref{eq:-8}) can alternatively be written as $\xi TD_{T}=X^*[x_\phi]$, where
\begin{equation}
x_\phi=h_{\phi}-\sum_{k\in \text{ions}} n_k^\phi\left( Q_{k0}^*+k_BT\right).\label{eq:-20}
\end{equation}
In aqueous suspensions, the interfacial excess enthalpy density resulting from charged colloid-fluid interactions is given by \cite{Burelbach2018b,Wurger2010}
\begin{equation}
h_\phi=\sum_{k\in \text{ions}}(n_kq_k\varphi+n_k^\phi k_BT)+\frac{1}{2}\epsilon_T\epsilon (\nabla\varphi)^2,
\end{equation}
where $\epsilon_T=\partial\ln\epsilon/\partial\ln T$. Substituting this into eq. (\ref{eq:-20}), we obtain 
\begin{equation}
x_\phi=\sum_{k\in \text{ions}} \left(n_kq_k\varphi-n_k^\phi Q_{k0}^*\right)+\frac{1}{2}\epsilon_T\epsilon (\nabla\varphi)^2.\label{eq:-16}
\end{equation}
The term $n_k q_k\varphi$ represents the electrostatic energy density of the double-layer ions of solute $k$. The contribution $-n_k^\phi Q_{k0}^*$ stems from the ionic steady-state condition (\ref{eq:-21}) and is related to the interfacial heat of ion hydration $Q_{k0}^*$. The last term in eq. (\ref{eq:-16}) represents the enthalpy density of the hydration layer around the charged colloid, assuming that the water molecules are freely polarisable at its surface \cite{Agar1989,Landau1960}. Using the quadratic form of $n_k^\phi$, eq. (\ref{eq:-16}) can be expressed as
\begin{equation}
x_\phi=-\epsilon\kappa^2\varphi^2+\frac{1}{2}\epsilon_T\epsilon(\nabla\varphi)^2-\epsilon\kappa^2\varphi(\varphi_T+\alpha\varphi),\label{eq:-3}
\end{equation}
where eq. (\ref{eq:-27}) has been used to identify $\varphi_T$. We have further introduced a new parameter $\alpha$ that quantifies the bulk gradient in ionic strength, defined as 
\begin{equation}
\alpha=\frac{1}{2k_BT}\frac{\sum_k n_k^bq_k^2Q_{k0}^*}{\sum_k n_k^bq_k^2}.\label{eq:-28}
\end{equation}
With eqs. (\ref{eq:-3}) and (\ref{eq:-11}), the interfacial transport coefficient $X^*[x_\phi]$ can finally be determined from eq. (\ref{eq:th-54}), giving
\begin{eqnarray}
\xi TD_T & = & 4\pi R\epsilon\zeta^2\kappa'^2\mathcal I^1_2\nonumber\\
& & -2\pi R\epsilon_T\epsilon\zeta^2\sum_{m=1}^3 \left(m\kappa'\right)^{3-m}\mathcal I_2^m\nonumber\\
& & +\xi\mu_E\varphi_T+4\pi R\alpha\epsilon\zeta^2\kappa'^2\mathcal I^1_2.\label{eq:-14}
\end{eqnarray}
The terms in eq. (\ref{eq:-14}) exactly correspond to the ones in eq. (\ref{eq:-3}). The first term is the electrostatic contribution, the second term accounts for colloid hydration, and the last two terms stem from ion hydration. The first term due to ion hydration is an electrophoretic contribution that scales with the electrophoretic mobility $\mu_E$, given by eq. (\ref{eq:-6}). The second term due to ion hydration is related to the salinity gradient $\alpha$ and can hence be interpreted as a diffusiophoretic contribution. Although the integrals $\mathcal I^m_2$ cannot be computed analytically for $m\geq1$, analytical expressions for $D_T$ can be obtained based on its limiting behaviour at low and high ionic strength. 

At low ionic strength (wide layers, $\kappa'\ll 1$), the surface potential $\zeta$ given by eq. (\ref{eq:-18}) is independent of $\kappa'$. In this case, the dominant terms in eq. (\ref{eq:-14}) are the electrophoretic contribution and the $\mathcal I_2^3$-term from colloid hydration. Moreover, the exponential screening in $\mathcal I_2^3$ can be neglected, such that $\mathcal I_2^3=\int_{1}^{\infty}\left(1-b\rho^{-1}\right)\rho^{-2}d\rho$. This yields $\mathcal I_2^3=1-b/2$, and the expression for the thermophoretic mobility becomes
\begin{equation}
D_T=-\frac{\epsilon_T\epsilon\zeta^2}{3b\eta T}\left(1-\frac{b}{2}\right)+\frac{\mu_E\varphi_T}{T},\hspace{0.4 cm}\kappa'\ll 1,\label{eq:-17}
\end{equation}
where we have used $\xi=6\pi b\eta R$. Equation (\ref{eq:-17}) differs from the result obtained in \cite{Morthomas2008} by a factor $(1-b/2)$ in the second term. The reason is that the work in \cite{Morthomas2008} considered the point-like limit ($R\rightarrow 0$) of eq. (\ref{eq:th-54}). Indeed, neglecting the term $-b\rho^{-1}$ in $\mathcal I_2^3$ just yields $\mathcal I_2^3=1$, without any further dependence on the slip coefficient $b$.

At high ionic strength (thin layers, $\kappa'\gg 1$), the surface potential $\zeta$ scales with $1/\kappa'$ and $\mathcal I^m_2$ can be evaluated analytically by performing a first-order expansion in the small parameter $\rho-1=(r-R)/R$, yielding $\mathcal I^m_2=\left(1+(2\kappa'-m)(1-b)\right)/(4\kappa'^2)$. The contributions can then be expressed to leading order in $\kappa'$, giving
\begin{eqnarray}
D_T & = & \frac{\epsilon\zeta^2}{6b\eta T}\left(b+2\kappa'(1-b)\right)\nonumber\\
& & -\frac{\epsilon_T\epsilon\zeta^2}{12b\eta T}\left(4-3b+2\kappa'(1-b)\right)\nonumber\\
& &+\frac{\mu_E\varphi_T}{T}+\frac{\alpha\epsilon\zeta^2}{6b\eta T}\left(b+2\kappa'(1-b)\right),\hspace{0.15 cm}\kappa'\gg 1.\nonumber
\end{eqnarray}
For weakly charged colloids, this expression coincides with the result obtained for a stick boundary ($b=1$) using a force-free approach based on the BLA \cite{Wurger2008,Wurger2010}, which successfully describes experimental data on polystyrene colloids \cite{Putnam2005}. The correspondence is obtained by noticing that the salinity gradient $\alpha$ in \cite{Wurger2008} is defined in terms of $Q_k^*$ rather than $Q_{k0}^*$ and differs from our definition (\ref{eq:-28}) by a factor 2. As the terms proportional to $\zeta^2$ always decay faster with $\kappa'$ than the one linear in $\zeta$, we further have 
\begin{equation}
D_T=\frac{\mu_E\varphi_T}{T},\hspace{0.4 cm}\kappa'\ggg 1.\label{eq:-25}
\end{equation}
Therefore, the electrophoretic contribution is expected to be a dominant contribution to thermophoresis of charged colloids at very high ionic strength.
\begin{figure}
	\centering{}\includegraphics[width=8.5cm,height=8cm]{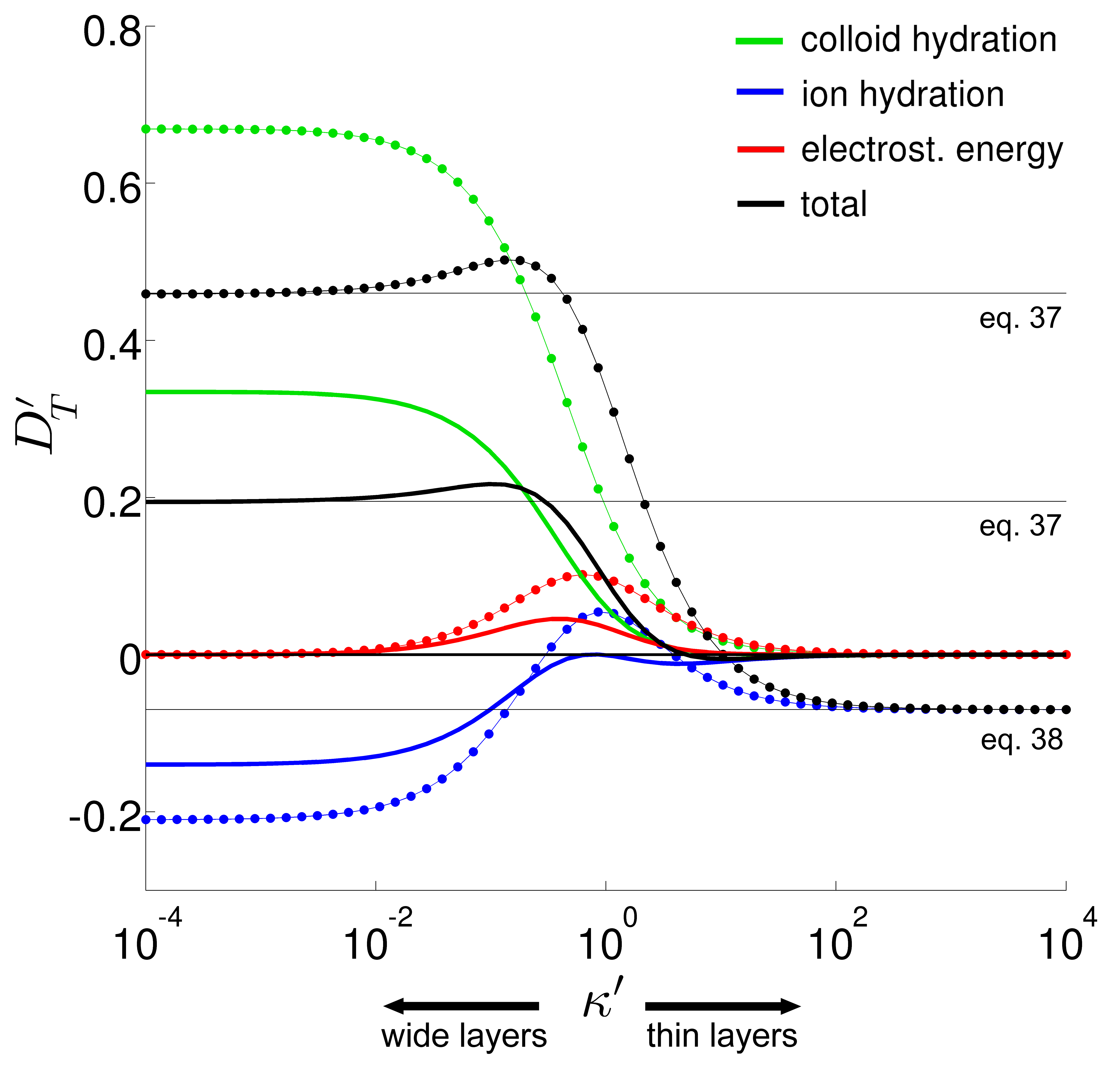}\caption{Rescaled  thermophoretic mobility $D_T'$ plotted vs. $\kappa'$, in the presence of NaOH. Each contribution is shown for a stick boundary (solid lines) and a perfect-slip boundary (dotted lines). The thin black lines show the corresponding limiting values.}\label{fig:-1}
\end{figure}

To study the general behaviour of the thermophoretic mobility in between the limiting cases, we solve the integrals in eq. (\ref{eq:-14}) numerically and reduce $D_T$ to a dimensionless function of $\kappa'$ and $b$, by applying the rescaling $D_T'=3\epsilon\eta T/(2\sigma^2 R^2)\times D_T$, such that: 
\begin{eqnarray}
D_T' & = & \frac{1}{b}\left(\frac{\kappa'}{1+\kappa'}\right)^2\mathcal I^1_2\nonumber\\
& & 
-\frac{\epsilon_T}{2b}\frac{\kappa'^2\mathcal I^1_2+2\kappa'\mathcal I^2_2+\mathcal I^3_2}{(1+\kappa')^2}\nonumber\\
& & +\frac{\varphi_T'}{b}\frac{1+\kappa'(1-b)}{(1+\kappa')}+\frac{\alpha}{b}\left(\frac{\kappa'}{1+\kappa'}\right)^2\mathcal I^1_2.\label{eq:-13}
\end{eqnarray}
Like the electrostatic contribution, the colloid hydration term is positive, as the quantity $\epsilon_T$ takes the value $-1.34$ for water at room temperature \cite{Dhont2008}. The remaining parameters that need to be specified are $\alpha$ and the ratio $\varphi_T'=\epsilon\varphi_T/(\sigma R)$ between the thermoelectric potential $\varphi_T$ and the bare colloidal surface potential $\sigma R/\epsilon$. Hence, only the ion hydration term can change its sign and magnitude relative to the other contributions. For common salts, acids and bases, the magnitude of $\varphi_T$ may reach up to $100\hspace{0.5mm}\text{mV}$ \cite{Wurger2010}, whereas the bare surface potential of colloids is of the order $\sim1$V \cite{Duhr2006b}. As an example, we consider a negatively charged colloid with a bare surface potential of $-0.5\hspace{0.5mm}\text{V}$ inside an aqueous solution exclusively titrated with the base NaOH ($\alpha = 2$, $\varphi_T=70\hspace{0.5mm}\text{mV}$ \cite{Wurger2010}). This parameter choice gives a positive diffusiophoretic contribution and a negative electrophoretic contribution, thus allowing a sign reversal of $D_T'$ at high ionic strength. 

Figure \ref{fig:-1} shows the rescaled thermophoretic mobility $D_T'$ and its different contributions plotted versus $\kappa'$, for both a stick boundary ($b=1$) and a perfect-slip boundary ($b=2/3$). For our parameter choice, the net mobility is positive at low ionic strength ($\kappa'\ll 1$) and negative at high ionic strength ($\kappa'\gg 1$), with a sign reversal occurring when $\kappa'> 1$. The colloid hydration term shows a strong enhancement for perfect slip and rapidly decreases with increasing ionic strength. The electrostatic contribution tends to zero in both limits, but reaches a peak in the intermediate regime $\kappa'\sim 1$. A more complicated trend is observed for the ion hydration term, as it comprises the electrophoretic and diffusiophoretic contribution, which are respectively related to the ionic bulk properties $\varphi_T$ and $\alpha$. The net mobility converges towards the electrophoretic contribution for $\kappa'\gg 1$ and does not vanish in this limit when $b\neq1$, as predicted by eqs. (\ref{eq:-25}) and (\ref{eq:-7}). In general, a complex thermophoretic behaviour should be expected in the intermediate regime $\kappa'\sim 1$, where all contributions have similar magnitudes.
\begin{figure}
	\centering{}\includegraphics[width=9cm,height=4.5cm]{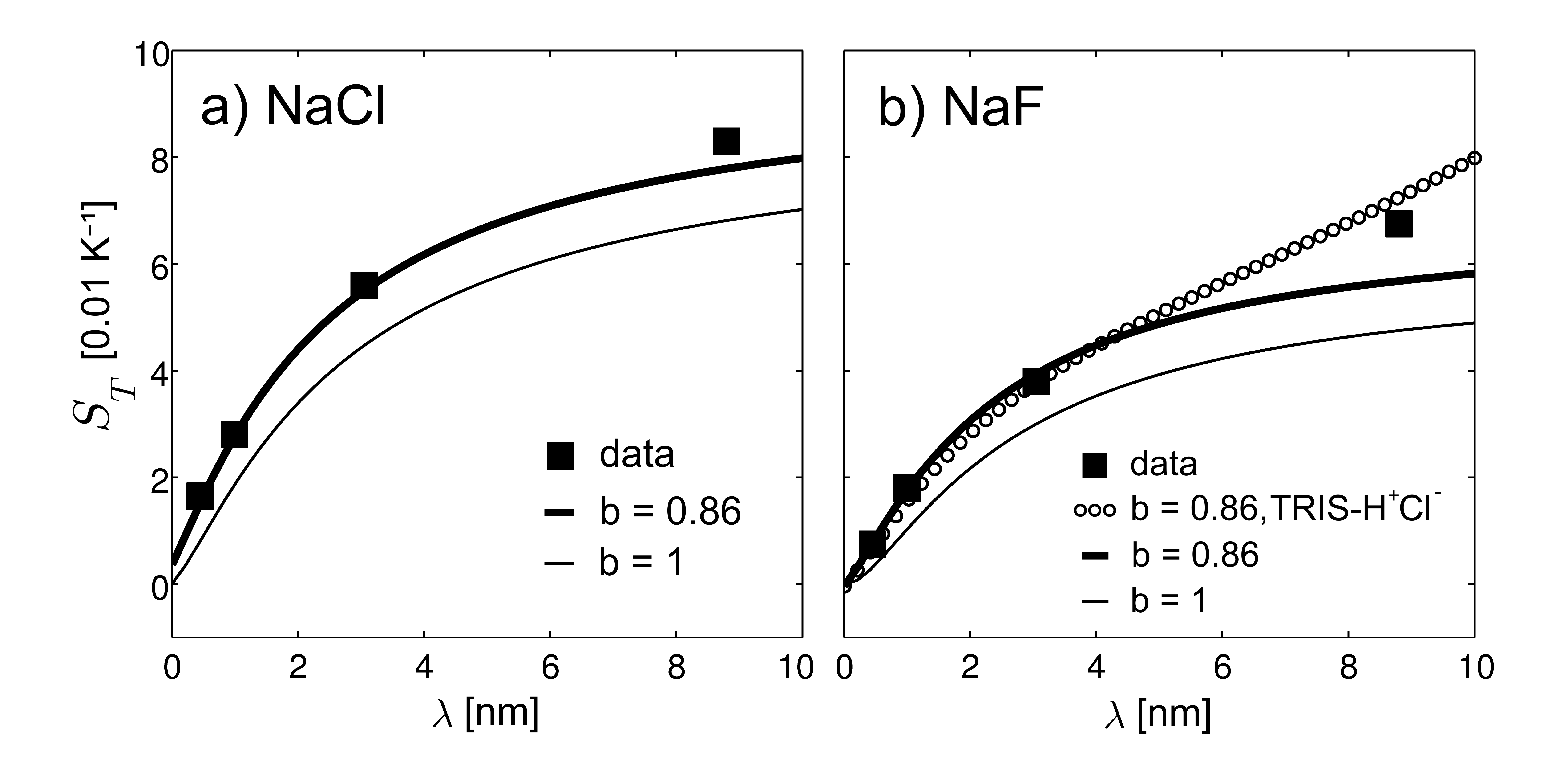}\caption{Soret coefficient $S_T$ vs. $\lambda$, for 22mer ssDNA titrated with a) NaCl and b) NaF. The data from \cite{Reichl2014} is fitted with $R=1.9\hspace{0.5mm}\text{nm}$ and an effective charge of $-12.3\hspace{0.5mm}e$, comparable to the values used in \cite{Reichl2014}. The values of $Q_{k0}^*$ for calculating $\varphi_T$ and $\alpha$ are also taken from \cite{Reichl2014}.}\label{fig:-2}
\end{figure}

We further compare eq. (\ref{eq:-14}) to experimental data on thermophoresis of DNA \cite{Reichl2014}. DNA has a negative charge and a hydrodynamic radius comparable to the Debye length ($R\sim \lambda$), making it an interesting candidate for the validation of our theory beyond the well-known limiting cases of wide and thin layers. Moreover, DNA has a polymeric structure that leaves the local gradients at its surface unperturbed, so that the assumption of uniform gradients is well-justified. In \cite{Reichl2014}, 22mer ssDNA was suspended in water containing $1\hspace{0.5mm}\text{mM}$ of TRIS-HCl buffer, followed by an addition of different salts. The measurements were described using the capacitor model \cite{Dhont2007,Dhont2008}, which evaluates the thermophoretic force from eq. (\ref{eq:-32}) and thereby ignores the hydrodynamic boundary condition at the particle surface. As the heat transport coefficient of TRIS-H$^+$ is not known, the effect of buffer dissociation was also ignored in \cite{Reichl2014}. However, TRIS-HCl is close to fully dissociated in the considered pH-range, thus setting an upper bound of about $10\hspace{0.5mm}\text{nm}$ for the Debye length at $1\hspace{0.5mm}\text{mM}$. Moreover, the electrophoretic contribution was fitted separately as a constant offset, even though $\mu_E$ is a well-defined function of ionic strength. Here, we provide an alternative fitting of this data based eq. (\ref{eq:-14}), by taking into account the dependence of $\mu_E$ on Debye length and by imposing a realistic hydrodynamic boundary condition at the DNA surface. 

Fig. \ref{fig:-2} shows the Soret coefficient $S_T=D_T/D$ versus the Debye length $\lambda$, in the presence of NaCl ($\varphi_T=-15\hspace{0.5mm}\text{mV}$, $\alpha=0.4$) and NaF ($\varphi_T=2\hspace{0.5mm}\text{mV}$, $\alpha=0.7$). The electrophoretic contribution of DNA is thus positive for NaCl, but negative for NaF. For NaCl (fig. \ref{fig:-2}a), the data is very well fitted by eq. (\ref{eq:-14}) if a partial slip is imposed at the DNA surface ($b=0.86$, thick line). A stick boundary ($b=1$, thin line) yields a similar trend, but is clearly lower in magnitude. The value $b=0.86$ is realistic for DNA as it corresponds to a slip length $l_s$ of $0.46\hspace{0.5mm}\text{nm}$, close to the value of $0.5\hspace{0.5mm}\text{nm}$ previously used to describe electrophoresis of DNA inside nanopores \cite{Galla2014}. As buffer dissociation has been ignored for the fitting in fig. \ref{fig:-2}a, the observed agreement suggests that TRIS-H$^+$ and Na$^+$ must have similar values of $Q_{k0}^*$, which is supported by measurements for similar organic compounds \cite{Takeyama1988}. This conclusion is also confirmed by fig. \ref{fig:-2}b, which shows the titration with NaF. Although a good agreement is observed at high ionic strength (small $\lambda$), the theoretical prediction (thick line) cannot explain the large value of $S_T$ at $\lambda\sim 9\hspace{0.5mm}\text{nm}$ if buffer dissociation is ignored. However, a very good fit is obtained if the heat of ion hydration $Q_{k0}^*$ of TRIS-H$^+$ is assumed to be the same as that of Na$^+$ (circles), showing that the ionic bulk properties $\varphi_T$ and $\alpha$ are indeed set by the dissociated buffer at low ionic strength.

\section{Conclusion}

We have used Onsager's reciprocal relations to determine the phoretic mobilities of weakly charged colloids. Our treatment generalises the results previously known for the cases of wide and thin interfacial layers and shows that phoretic motion is sensitive to the hydrodynamic boundary condition at the colloidal surface. In particular, we have shown that a slip boundary condition leads to a non-vanishing electrophoretic mobility at high ionic strength. Our expression for the thermophoretic mobility is in agreement with existing limiting results and successfully describes experimental data on DNA. Our comparison to these experiments also suggests that buffer dissociation matters at low ionic strength and that DNA has a slippery surface. The latter conclusion has previously been drawn for electrophoresis of DNA and highlights the hydrodynamic character of phoretic motion.


\bibliographystyle{acm}
\bibliography{References_EPJE}

\begin{thebibliography}{10}

\bibitem{Agar1989}
{\sc Agar, J.~N., Mou, C.~Y., and Lin, J.~L.}
\newblock {\em J. Phys. Chem. 93\/} (1989), 2079--2082.

\bibitem{Ajdari2006}
{\sc Ajdari, A., and Bocquet, L.}
\newblock {\em Phys. Rev. Lett. 96\/} (2006), 186102.

\bibitem{Anderson1989}
{\sc Anderson, J.~L.}
\newblock {\em Annu. Rev. Fluid Mech. 21\/} (1989), 61--99.

\bibitem{Barber}
{\sc Barber, R., and Emerson, D.}
\newblock {\em Analytical solution of low Reynolds number slip flow past a
  sphere}.
\newblock Council for the Central Laboratory of the Research Councils, 2000.

\bibitem{Brady2011}
{\sc Brady, J.~F.}
\newblock {\em J. Fluid Mech. 667\/} (2011), 216--259.

\bibitem{Burelbachthesis}
{\sc Burelbach, J.}
\newblock PhD thesis, University of Cambridge, 2018.

\bibitem{Burelbach2018}
{\sc Burelbach, J., Br{\"{u}}ckner, D.~B., Frenkel, D., and Eiser, E.}
\newblock {\em Soft Matter 14\/} (2018), 7446--7454.

\bibitem{Burelbach2018b}
{\sc Burelbach, J., Frenkel, D., Pagonabarraga, I., and Eiser, E.}
\newblock {\em Eur. Phys. J. E 41\/} (2018), 7.

\bibitem{Derjaguin1987}
{\sc Churaev, N.~V., Derjaguin, B.~V., and Muller, V.~M.}
\newblock {\em Surface forces}.
\newblock Consultants Bureau, New York, 1987.

\bibitem{deGroot1962}
{\sc De~Groot, S.~R., and Mazur, P.}
\newblock {\em Non-equilibrium thermodynamics}.
\newblock Courier Corporation, 2013.

\bibitem{Derjaguin1966}
{\sc Derjaguin, B., and Yalamov, Y.}
\newblock {\em J. Colloid Sci. 20\/} (1965), 555--570.

\bibitem{Dhont2007}
{\sc Dhont, J.~K., Wiegand, S., Duhr, S., and Braun, D.}
\newblock {\em Langmuir 23\/} (2007), 1674--1683.

\bibitem{Dhont2008}
{\sc Dhont, J. K.~G., and Briels, W.~J.}
\newblock {\em Eur. Phys. J. E 25\/} (2008), 61--76.

\bibitem{Duhr2006b}
{\sc Duhr, S., and Braun, D.}
\newblock {\em Proceedings of the National Academy of Sciences of the United
  States of America 2006\/} (2006), 19678–19682.

\bibitem{Fayolle2008}
{\sc Fayolle, S., Bickel, T., and W\"{u}rger, A.}
\newblock {\em Phys. Rev. E 77\/} (2008), 041404.

\bibitem{Galla2014}
{\sc Galla, L., Meyer, A.~J., Spiering, A., Sischka, A., Mayer, M., Hall,
  A.~R., Reimann, P., and Anselmetti, D.}
\newblock {\em Nano Lett. 14\/} (2014), 4176--4182.

\bibitem{Gaspard2018}
{\sc Gaspard, P., and Kapral, R.}
\newblock {\em J. Chem. Phys. 148\/} (2018), 134104.

\bibitem{huckel1924cataphoresis}
{\sc H{\"u}ckel, E.}
\newblock {\em Phys. Zeitschrift 25\/} (1924), 204--210.

\bibitem{Khair2009}
{\sc Khair, A.~S., and Squires, T.~M.}
\newblock {\em Phys. Fluids 21\/} (2009), 042001.

\bibitem{Landau1987}
{\sc Landau, L., and Lifshitz, E.}
\newblock {\em Fluid mechanics}.
\newblock London, 1959.

\bibitem{Landau1960}
{\sc Landau, L.~D., Bell, J., Kearsley, M., Pitaevskii, L., Lifshitz, E., and
  Sykes, J.}
\newblock {\em Electrodynamics of continuous media}, vol.~8.
\newblock elsevier, 2013.

\bibitem{Morthomas2008}
{\sc Morthomas, J., and W\"urger, A.}
\newblock {\em Eur. Phys. J. E 27\/} (2008), 425.

\bibitem{Morthomas2009}
{\sc Morthomas, J., and W{\"{u}}rger, A.}
\newblock {\em J. Phys.: Condens. Matter 21\/} (2009), 035103.

\bibitem{Onsager1931}
{\sc Onsager, L.}
\newblock {\em Phys. Rev. Lett. 37\/} (1931), 405.

\bibitem{Onsager1931a}
{\sc Onsager, L.}
\newblock {\em Phys. Rev. Lett. 38\/} (1931), 2265.

\bibitem{Parola2004}
{\sc Parola, A., and Piazza, R.}
\newblock {\em Eur. Phys. J. E 15\/} (2004), 255--63.

\bibitem{Piazza2004}
{\sc Piazza, R.}
\newblock {\em J. Phys.: Condens. Matter 16\/} (2004), S4195--S4211.

\bibitem{Piazza2008}
{\sc Piazza, R., and Parola, A.}
\newblock {\em J. Phys.: Condens. Matter 20\/} (2008), 153102.

\bibitem{Putnam2005}
{\sc Putnam, S.~A., and Cahill, D.~G.}
\newblock {\em Langmuir 21\/} (2005), 5317--5323.

\bibitem{Reichl2014}
{\sc Reichl, M.}
\newblock PhD thesis, Ludwig Maximilian University of Munich, 2014.

\bibitem{Ruckenstein1981}
{\sc Ruckenstein, E.}
\newblock {\em J. Coll. Int. Sci. 83\/} (1981), 77--81.

\bibitem{Semenov2015}
{\sc Semenov, S., and Schimpf, M.}
\newblock {\em J. Phys. Chem. B 119\/} (2015), 3510--3516.

\bibitem{smoluchowski1903contribution}
{\sc Smoluchowski, M.~v.}
\newblock {\em Bull Int Acad Sci Cracovie 3\/} (1903), 184--199.

\bibitem{Takeyama1988}
{\sc Takeyama, N., and Nakashima, K.}
\newblock {\em J. Sol. Chem. 17\/} (1988), 305--325.

\bibitem{Wurger2007}
{\sc W{\"{u}}rger, A.}
\newblock {\em Phys. Rev. Lett. 98\/} (2007), 138301.

\bibitem{Wurger2008}
{\sc W{\"{u}}rger, A.}
\newblock {\em Phys. Rev. Lett. 101\/} (2008), 108302.

\bibitem{Wurger2010}
{\sc W\"{u}rger, A.}
\newblock {\em Rep. Prog. Phys. 73\/} (2010), 126601.

\end{thebibliography}

\end{document}